\documentclass[twocolumn,secnumarabic,amssymb, amsmath, nobibnotes, aps, prd]{revtex4-1}

\usepackage[dvipdfmx]{graphicx}  
\usepackage[dvipdfmx]{color}
\usepackage{dcolumn}   
\usepackage{bm}        
\usepackage{amssymb}   
\usepackage{here}
\usepackage{amsmath}
\usepackage{comment}
\usepackage{color}

\usepackage{braket}
\usepackage{mathrsfs}
\usepackage{amsmath}
\usepackage{ulem}
\usepackage{fancybox}
\usepackage{xspace}
\usepackage{amsfonts}
\usepackage{multirow}
\usepackage{amsthm}
\usepackage{setspace}
\usepackage[subrefformat=parens]{subcaption}
\usepackage{ulem}
\usepackage[version=3]{mhchem}

\begin{document}

\title{Dirac surface states in magnonic analogs of topological crystalline insulators}%

\author{Hiroki Kondo and Yutaka Akagi}
\affiliation{
Department of Physics, Graduate School of Science, The University of Tokyo, 7-3-1 Hongo, Tokyo 113-0033, Japan}
\email[]{kondo-hiroki290@g.ecc.u-tokyo.ac.jp}

\begin{abstract}
We propose magnonic analogs of topological crystalline insulators which possess Dirac surface states protected by the combined symmetry of time-reversal and half translation.
Constructing models of the topological magnon systems, we demonstrate that an energy current flows through the systems in response to an electric field,
owing to the Dirac surface states with the spin-momentum locking.
We also propose a realization of the magnonic analogs of topological crystalline insulators in a magnetic compound CrI$_{3}$ 
with a monoclinic structure.
\end{abstract}

\maketitle

{\it Introduction.}---
There has been an explosion of interest in the topological properties of condensed matter systems~\cite{Klitzing80,Thouless82,Kohmoto85,Schnyder08,Kitaev09,Ryu10,Hasan10,Qi11}, 
especially since the proposal of topological insulators (TI)~\cite{Kane05a,Kane05b,Fu07} 
robust against perturbations such as disorder~\cite{Sheng-Haldane06,Nomura07,Obuse07,Essin07,Jiang09,Li09,Groth09,Guo10a,Loring10,Prodan11,Yamakage11,Fulga12,Leung12,Kobayashi13,Sbierski14,Katsura16,Akagi17,Katsura18}.
One of their hallmarks is the appearance of the Dirac surface states protected by time-reversal symmetry.
During the last decade, it has been recognized that crystalline symmetries lead to a more refined classification of phases, so-called topological crystalline insulators (TCI)~\cite{Fu11,Tanaka12,Dziawa12,Shiozaki14}.
Among them, of particular interest are antiferromagnetic topological insulators~\cite{Mong10}, known as one of the earliest proposals of TCI.
They can be roughly regarded as a stack of quantum  Hall insulators with alternating Chern numbers.
They have Dirac surface states topologically protected by the combined symmetry of time-reversal and translation of half a unit cell in the stacking direction.

Topological phases and phenomena have also been explored intensively in bosonic systems
such as systems of magnons~\cite{Fujimoto09, Katsura10, Matsumoto11a, Matsumoto14, Shindou13a, Shindou13b,Kim16, Onose10, Ideue12,Chisnell15, Han_Lee17, Murakami_Okamoto17,Kawano19a,Li18,Hirschberger15,Li17,Pershoguba18,Bao18}, 
photons~\cite{Onoda04, Hosten08, Raghu08, Haldane08, Wang09a, Ben-Abdallah16,Lu13}, 
phonons~\cite{Strohm05, Sheng06, Inyushkin07, Kagan08, Wang09b, Zhang10, Qin12, Huber16, Sugii17, Mori17, Susstrunk15,Stenull16},
triplons~\cite{Rumhanyi15, McClarty17, Joshi17, Joshi19, Nawa19},
and Bose-Einstein condensates~\cite{Furukawa15, Engelhardt15, Bardyn16, Xu-You16, Liberto16, Pan16, Luo18, Yoshino19, Ohashi20}.
The topological classification is more complicated than that in fermionic systems because bosonic Bogoliubov--de Gennes (BdG) systems possess non-Hermicity due to Bose statistics~\cite{Lieu18, Kawabata19, Kondo20}.
Meanwhile, topological bosonic systems can exhibit fascinating transport phenomena which are qualitatively different from those in fermionic systems.
For example, in the magnonic Weyl semimetals~\cite{Li16,Mook16,Su17,Liu19,Owerre18},
magnons having no electric charge can be driven by an electric field due to the chiral anomaly.
As examples of symmetry-protected topological phases for bosons,
magnonic analogs of quantum spin Hall insulators~\cite{Zyuzin16,Nakata17,Kondo19a} and three-dimensional TI~\cite{Kondo19b} were proposed recently.
However, the latter system possessing a single surface Dirac state is quite artificial; i.e., the system is a kind of a ``bilayer" diamond lattice which is constructed so that they have pseudo-time-reversal symmetry~\cite{comment_artificial}.
Hence, its realization in real materials seems to be difficult. 


In this Letter, 
we construct a model of a three-dimensional magnet which has single magnon surface states protected by the symmetry of the combined operation of time-reversal ($\Theta$) and half translation ($T_{1/2}$).
We represent the combined operator as 
$S=\Theta T_{1/2}$ and  refer to the symmetry as $S$-symmetry.
The model describes a magnonic analog of antiferromagnetic topological insulators (MAFTI)~\cite{Mong10} or TCI.
Since magnons are bosons, the time-reversal operator squares to the identity,
which does not ensure the existence of Kramers pairs.
On the other hand, $S$-symmetry leads to Kramers degeneracy at certain wave vectors, at which topologically protected surface Dirac states can exist.

In addition, we show that a homogeneous electric field induces the imbalance of the position of the surface Dirac cones
between opposite surfaces in MAFTI, which gives rise to an energy current.
Owing to the Aharonov-Casher (AC) effect~\cite{Aharonov84}
and the spin-momentum locking~\cite{Okuma17,Kawano19b,Kawano19c} in the magnon surface states, 
the system exhibits such an intrinsic electric-field response whose counterpart is absent in electronic systems.
We evaluate the energy current driven by an electric field by using linear response theory.
As an advantage of the concept for MAFTI, we also propose that the magnetic compound CrI$_{3}$~\cite{McGuire15,Huang17,Huang18,Chen18,Sivadas18,Soriano19,Ubrig19,Costa20,Soriano20} with a monoclinic structure is a candidate material for MAFTI.

\begin{figure}[H]
\centering
  \includegraphics[width=5cm]{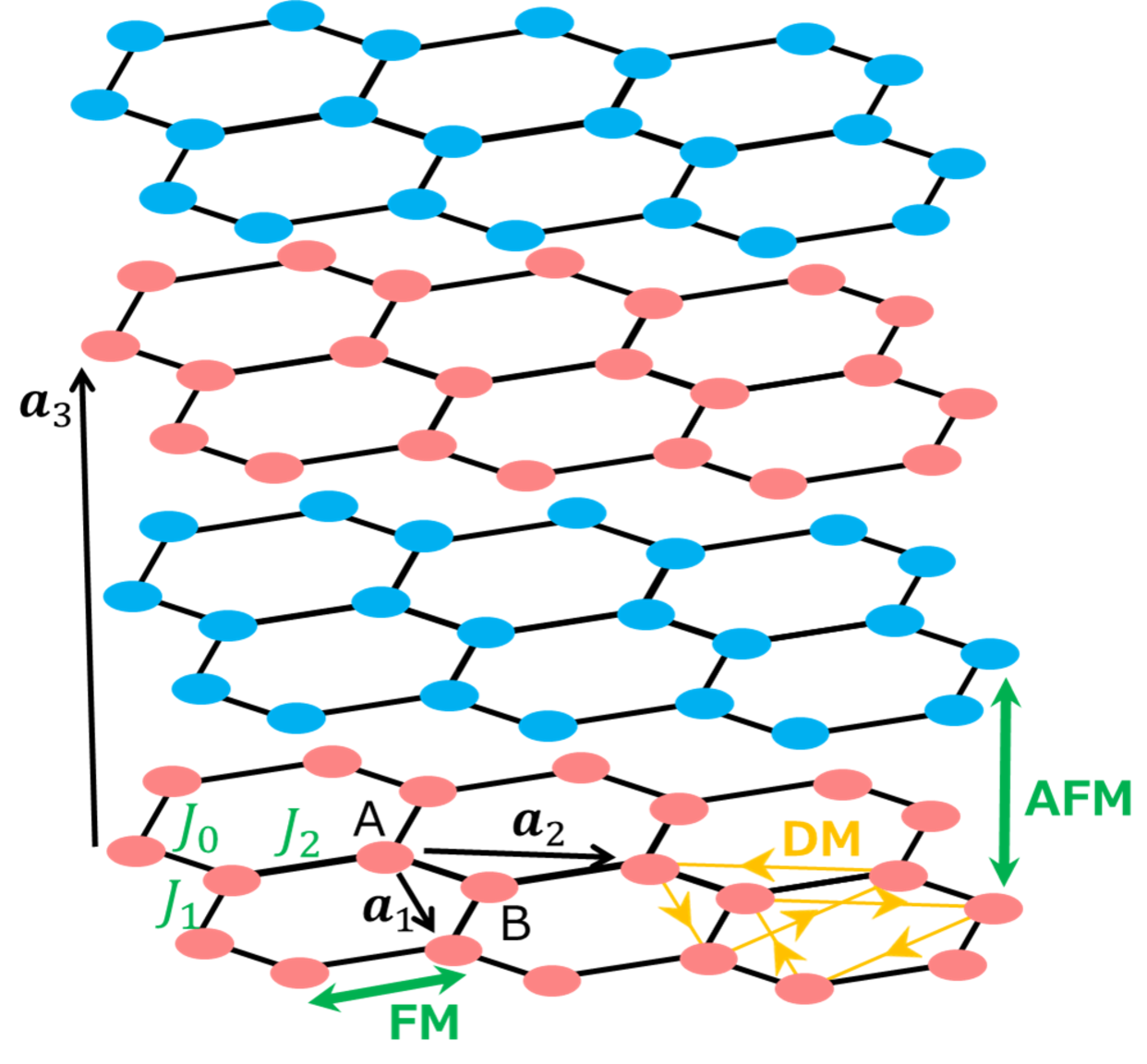}
\caption{(color online). Stacked honeycomb lattice magnet. 
Red and blue circles represent spins pointing in the $+z$- and $-z$-directions, respectively.
The two sublattices of each layer are indicated by A and B.
The three bond-dependent couplings are represented by $J_n$ ($n=0$, $1$, $2$).
The orange arrows indicate the sign convention $\xi_{ij}=+1(=-\xi_{ji})$ for $i\to j$. 
The vectors $\bm{a}_{1}$, $\bm{a}_{2}$, and $\bm{a}_{3}$ are the lattice primitive vectors of the lattice. 
}\label{fig:HoneycombTCI_AA}
\end{figure}

{\it Models.}---
As a model for MAFTI, we consider a stack of honeycomb lattice magnets with intralayer ferromagnetic and interlayer antiferromagnetic interactions,
where the spins on the same layer (odd and even layers) are aligned in the same direction (opposite directions) [see Fig.~\ref{fig:HoneycombTCI_AA}].
The Hamiltonian is given as follows:
\begin{align}
\mathcal{H}=
&-\sum_{\langle ij \rangle,l}\bm{S}_{i,l} J_{ij}\bm{S}_{j,l}+D\sum_{\langle\langle ij \rangle\rangle,l}\xi_{ij}\left(\bm{S}_{i,l}\times\bm{S}_{j,l}\right)_{z} \nonumber \\
&+J'\sum_{i,\langle l,l' \rangle}\bm{S}_{i,l}\cdot\bm{S}_{i,l'}.
\label{eq:Ham_AA}
\end{align}
Here, 
$\bm{S}_{j,l}$ is given by $\bm{S}_{j,l} := (S_{j,l}^x, S_{j,l}^y, S_{j,l}^z)$, where
$S_{j,l}^{\gamma}$ is $\gamma$-component ($\gamma=x$, $y$, $z$) of the spin operator.
The subscripts $i,j$ and $l,l'$ are the labels for the sites in honeycomb lattices and for the layers, respectively.
Here, the first term is the nearest neighbor ferromagnetic Heisenberg interaction with bond dependence and XYZ anisotropy.
The bond-dependent matrix $J_{ij}$ is a 3 by 3 diagonal matrix $J_{ij}:=J_{n}={\rm diag}(J^{x}_{n},J^{y}_{n},J^{z}_{n})$ $(n=0,1,2)$ for the three different bonds $\langle ij \rangle$ in the honeycomb lattice shown in Fig.~\ref{fig:HoneycombTCI_AA}. 
The second term represents the Dzyaloshinskii-Moriya (DM) interaction between next-nearest neighbor sites,
where $\xi_{ij}$ is a sign convention described by orange arrows in Fig.~\ref{fig:HoneycombTCI_AA}.
The remaining term is the antiferromagnetic Heisenberg interaction between the nearest neighbor layers.

By applying Holstein-Primakoff~\cite{Holstein40} and Fourier transformations, we can rewrite the Hamiltonian~(\ref{eq:Ham_AA}) as
\begin{align}
\mathcal{H}
&=\frac{1}{2}\sum_{\bm{k}}[\bm{b}^{\dagger}(\bm{k})\bm{b}(-\bm{k})]H(\bm{k})
\left[
\begin{array}{cc}
\bm{b}(\bm{k})  \\
\bm{b}^{\dagger}(-\bm{k})  \\
\end{array} 
\right],
\label{eq:Ham}
\end{align}
where $\bm{b}(\bm{k})=(b(\bm{k},A,1),b(\bm{k},B,1),b(\bm{k},A,2),b(\bm{k},B,2))^{T}$.
The operator $b(\bm{k},A(B),1(2))$ annihilates a magnon at the sublattice $A(B)$ on the layer with odd (even) $l$.
The matrix $H(\bm{k})$ is given in Supplemental Materials~\cite{comment_Suppl}.

The Hamiltonian matrix has $S$-symmetry: $S^{-1}(k_{z})H(\bm{k})S(k_{z})=H(-\bm{k})$, 
where $S(k_{z})$ is the combination operator defined by 
$S(k_{z})= \Theta T_{1/2}(k_{z})$.
The time-reversal operator $\Theta$ and the translation of half a unit cell in the 
$z$-direction $T_{1/2}(k_{z})$ are defined as 
$\Theta=K$ and
$T_{1/2}(k_{z})=1_{2}\otimes \sigma_x {\rm diag}(1,e^{ik_{z}})\otimes 1_{2}$ which satisfies $T_{1/2}(k_{z})^2=e^{ik_{z}}$, respectively.
Here $K$, $1_{2}$, and $\sigma_{\gamma}$ $(\gamma=x,y,z)$, are the complex conjugation, the 2 by 2 identity matrix, and the Pauli matrices, respectively.

\begin{figure}[H]
\centering
  \includegraphics[width=8.5cm]{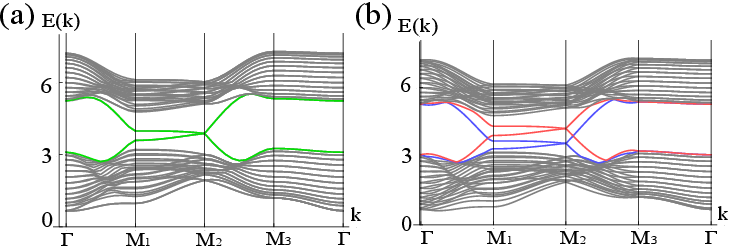}
\caption{(color online). (a) Magnon band structure of a slab with $(100)$ face of stacked honeycomb lattice magnet, 
and (b) that under an electric field in the $x$-direction.
Topologically protected surface Dirac states are shown
in (a) green and (b) red/blue.
The parameters 
are $J^{x}_{1}=1.6$, $J^{y}_{1}=0.4$, 
$J^{x}_{2}=1.0$, $J^{y}_{2}=1.0$, 
$J^{x}_{3}=0.4$, $J^{y}_{3}=1.6$, $J^{z}_{1}=J^{z}_{2}=J^{z}_{3}=1.1$, $D=0.2$, $J'=0.5$, and $S=1.0$.
Taking $\bm{a}_{1}=(1,0,0)$, $\bm{a}_{2}=(0,1,0)$ and $\bm{a}_{3}=(0,0,1)$, we here deform the stacked honeycomb lattice into a topologically equivalent cubic-shaped lattice~\cite{comment_Suppl}.
The symmetry points are $\Gamma=(0,0)$, ${\rm M}_{1}=(\pi,0)$, ${\rm M}_{2}=(\pi,\pi)$, ${\rm M}_{3}=(0,\pi)$.
}\label{fig:HoneycombTCI_EF}
\end{figure}

The equation $S(-k_{z})S(k_{z})=e^{i k_{z}}$ 
yields $S^{2}(\pi)=-1$, which 
leads to ${\mathbb Z}_{2}$ topological characterization 
in the same way as the magnonic analog of quantum spin Hall insulators~\cite{Kondo19a}.
Due to the Kramers theorem and the above relation, 
the bands of the model are doubly degenerate at time-reversal invariant momenta (TRIM) in $k_{z}=\pi$ plane. 
As in Fig.~\ref{fig:HoneycombTCI_EF}(a) which shows the band structure of a slab with $(100)$ face, 
a single Dirac cone can be found at $(k_{y},k_{z})=(\pi,\pi)$ (${\rm M}_{2}$ point). 
The situation is similar to that in strong topological insulators in class AII in the sense that a single Dirac cone appear.
Then, the corresponding ${\mathbb Z}_{2}$ topological invariant of MAFTI is defined as
\begin{align}
\nu_{z,\pi}^{n\sigma} \!\! := \!\!\frac{1}{2\pi} \! \!\left[ \! \oint_{\partial {\rm EBZ}_{z,\pi}} \!\!\!\!\!\!\!\!\!\!\!\!\!\!\!\!\!\! d\bm{k} \!\cdot \!\left[\bm{A}_{n\sigma}(\bm{k})\right]_{k_{z}=\pi} \!\!\! - \!\!\! \int_{{\rm EBZ}_{z,\pi}} \!\!\!\!\!\!\!\!\!\!\!\!\!\!\!\! dk_{x}dk_{y} \! \left[\Omega_{n\sigma}^{z}(\bm{k})\right]_{k_{z}=\pi} \! \right]\hspace{0mm}{\rm mod}\hspace{1mm}2.
\label{eq:Z2_MAFTI}
\end{align}
Here, we obtained Eq.~(\ref{eq:Z2_MAFTI}) by replacing the pseudo-time-reversal operator $\Theta'$ with $S(\pi)$ in the definition of $\nu^{n\sigma}_{z,\pi}$ in Ref.~\cite{Kondo19b}.
We also confirmed the correspondence between the existence (absence) of Dirac surface states and $\nu_{z,\pi}^{n\sigma}=1$ ($0$) by constructing the phase diagram of the model~(\ref{eq:Ham_AA}).
We note that the first, second, and third terms of the Hamiltonian in Eq.~(\ref{eq:Ham_AA}) are all necessary~\cite{comment_role} 
to realize the band structure having a single Dirac cone as in Fig.~\ref{fig:HoneycombTCI_EF}(a).


{\it Energy current induced by a homogeneous electric field.}---
Next, we propose an intrinsic field response in MAFTI, 
analogous to the topological magnetoelectric effect~\cite{Qi-Zhang08,Nomura11,Morimoto15}
for electrons, whereas the mechanism is essentially different from that in electronic systems.
Figure~\ref{fig:HoneycombTCI_EF}(b) represents the band structure of the model under an electric field $\bm{E}=E_{x}\bm{e}_{x}$.
As shown in the figure, the electric field shifts dispersions on the surface states upward (red) and downward (blue). 
We will show that the imbalance of the dispersions of the surface states results in an energy current. 
This is an intrinsic phenomenon realized by the spin-momentum locking of the magnon surface states~\cite{Okuma17,Kawano19b,Kawano19c}.

To understand the phenomena, let us consider the AC effect, in which magnons acquire a geometric phase by moving in an electric field $\bm{E}$.
By applying an electric field, the vector potential $-(\sigma g\mu_{\rm B}/c^{2})\bm{E}\times\bm{e}_{z}$ is added to the wave vector of magnons:
\begin{align}
\bm{k}\to \bm{k} -\frac{\sigma g\mu_{\rm B}}{c^{2}}\bm{E}\times\bm{e}_{z}.
\label{eq:AC}
\end{align}
Here, $c$ is the speed of light in a vacuum.
The vector 
$- \sigma g \mu_{\rm B}\bm{e}_{z} (:=\bm{\mu})$
is the magnetic moment of magnons from up ($\sigma=+$) or down ($\sigma=-$) spins, where
$g$ is the $g$-factor of the spins, $\mu_{\rm B}$ is the Bohr magneton, and
$\bm{e}_{\gamma}$ ($\gamma=x$, $y$, $z$) is the unit vector in the $\gamma$-direction.

Next, we show that the shift of the wave vector expressed by Eq.~(\ref{eq:AC}) gives rise to the shift of surface Dirac dispersions as in Fig.~\ref{fig:HoneycombTCI_EF}(b).
We write the effective Hamiltonian with spin-momentum locking for $(\bar{1}00)$ and $(100)$ surfaces as $H_{+}(\bar{\bm{k}})$ and $H_{-}(\bar{\bm{k}})$, respectively. Here, $\bar{\bm{k}}$ is defined as $\bar{\bm{k}}=(k_{y},k_{z})$.
They 
can be written as follows:
\begin{align}
H_{\pm}(\bar{\bm{k}})=\pm\left(
\begin{array}{cc}
\alpha k_{y} & \beta^{*} k_{z}  \\
\beta k_{z} & -\alpha k_{y}   \\
\end{array} 
\right),
\end{align}
where 
the coefficients $\alpha$ and $\beta$ are determined numerically (see Ref.~\cite{Kondo19b} for the derivation).
We note that the magnon state expressed by the wave function $\bm{\psi}=(1,0)^{T}$ ($\bm{\psi}'=(0,1)^{T}$) has 
the magnetic moment $\bm{\mu}=-g \mu_{\rm B}\bm{e}_{z}$ 
($\bm{\mu}=+g \mu_{\rm B}\bm{e}_{z}$). 
The sign $+/-$ of $H_{\pm}(\bar{\bm{k}})$ corresponds to the chirality of the magnon Dirac state.
By the Peierls substitution in Eq.~(\ref{eq:AC}), the effective Hamiltonian 
under a homogeneous electric field in the $x$-direction $\bm{E}=E_{x}\bm{e}_{x}$ can be written as follows:
\begin{align}
H_{\pm}(\bar{\bm{k}})
&\to \pm \left(
\begin{array}{cc}
\alpha (k_{y}+\frac{g\mu_{\rm B}}{c^{2}}E_{x}) & \beta^{*} k_{z}  \\
\beta k_{z} & -\alpha (k_{y}-\frac{g\mu_{\rm B}}{c^{2}}E_{x})   \\
\end{array} 
\right) \nonumber \\
&=H_{\pm}(\bar{\bm{k}})
\pm\alpha\frac{g\mu_{\rm B}}{c^{2}}E_{x}1_{2}.
\end{align}
Therefore, the energy of the gapless point of the Dirac state on $(\bar{1}00)$ surface shifts by $+\alpha(g\mu_{\rm B}/c^{2})E_{x}$, while the other shifts by $-\alpha(g\mu_{\rm B}/c^{2})E_{x}$.

To see the effect on these shifts, we here introduce the energy current operator defined as~\cite{Paul03,Matsumoto14}:
\begin{align}
J_{x}=\sum_{i}\bar{P}_{i}\dot{h}_{i},
\end{align}
where $\bar{P}_{i}$ and $\dot{h}_{i}$ are the position along the 
$x$-direction and the time derivative of the Hamiltonian density at the site $i$, respectively.
It can be rewritten as 
\begin{align}
J_{x}=\frac{1}{2}\sum_{\bar{\bm{k}}}\bm{\psi}^{\dagger}(\bar{\bm{k}})J_{x}(\bar{\bm{k}})\bm{\psi}(\bar{\bm{k}}),
\end{align}
where   $\bm{\psi}(\bar{\bm{k}})$ is a set of magnon operators.
The matrix $J_{x}(\bar{\bm{k}})$  is given by
\begin{align}
J_{x}(\bar{\bm{k}})  =-\frac{i}{2}\left(\bar{P}H(\bar{\bm{k}})\Sigma_{z}H(\bar{\bm{k}})
-H(\bar{\bm{k}})\Sigma_{z}H(\bar{\bm{k}})\bar{P}\right),
\end{align}
where $H(\bar{\bm{k}})$ and $\bar{P}$ are the matrix forms of the Hamiltonian
for a slab geometry with open boundary condition in the $x$-direction
and the position operator in the $x$-direction, respectively. 
The matrix $\Sigma_{z}$ is defined as $\Sigma_{z}=\sigma_{z}\otimes 1_{4N}$, where $1_{4N}$ is the $4N \times 4N$ identity matrix, and $N$ is the number of the unit cells in the $x$-direction parallel to the vector ${\bm a}_1$.
The details of the energy current operator are 
given in Supplemental Materials~\cite{comment_Suppl}.

To evaluate the energy current induced by an electric field, we use the linear response theory, considering the field $E_{x}(t)=E_{x}e^{-i\omega t}$.
Here, we describe the unperturbed Hamiltonian and the perturbation as $H_{0}$ and $H_{E}(t)=H_{E}e^{-i\omega t}$, respectively. 
The perturbation $H_{E}$ is the Hamiltonian of the first order in the electric field $E_{x}$ ~\cite{comment_Suppl}.
The expectation value of the energy current in the $x$-direction $J_{x}$ is given as follows:
\begin{align}
&\langle J_{x} \rangle \nonumber \\
&=-\frac{i}{\hbar}\int_{0}^{\infty}d\tau e^{i\omega \tau}{\rm tr}\left[
e^{-i H_{0}\tau/\hbar}[H_{E},\rho_{0}]e^{i H_{0}\tau/\hbar}J_{x}
\right] e^{-i\omega t}, 
\end{align}
where $\rho_{0}$ is the density operator for $H_{0}$ at thermal equilibrium.
Hereafter, we use Planck units, i.e.,
the Planck constant $\hbar=1$, the Boltzmann constant $k_{\rm B}=1$, and $c=1$.
In the limit of $\omega \to 0$, 
$\langle J_{x} \rangle$ is written as follows:
\begin{align}
&\langle J_{x} \rangle \nonumber \\
&=i\sum_{\bar{\bm{k}}}\sum_{\alpha \beta \gamma \delta \zeta \eta} \frac{n_{\rm B}(E_{\alpha}(\bar{\bm{k}}))-n_{\rm B}(E_{\delta}(\bar{\bm{k}}))}{(E_{\alpha}(\bar{\bm{k}})-E_{\delta}(\bar{\bm{k}})+i/\tau_{\ell})(E_{\alpha}(\bar{\bm{k}})-E_{\delta}(\bar{\bm{k}}))}\nonumber \\
&\times T_{\alpha \beta}^{-1}(\bar{\bm{k}})\left(\hat{J}_{x}(\bar{\bm{k}})\right)_{\beta \gamma}T_{\gamma \delta}(\bar{\bm{k}})T_{\delta \zeta}^{-1}(\bar{\bm{k}})\left(\hat{\dot{H}}_{E}(\bar{\bm{k}})\right)_{\zeta \eta}T_{\eta \alpha}(\bar{\bm{k}}),
\label{eq:energy_current}
\end{align}
where $n_{\rm B}$ is the Bose distribution function.
The matrices $\hat{J}_{x}(\bar{\bm{k}})$ and 
$\hat{\dot{H}}_{E}(\bar{\bm{k}})$ are defined 
as $\hat{J}_{x}(\bar{\bm{k}})=\Sigma_{z}J_{x}(\bar{\bm{k}})$ and $\hat{\dot{H}}_{E}(\bar{\bm{k}})=i(\Sigma_{z}H(\bar{\bm{k}})\Sigma_{z}H_{E}(\bar{\bm{k}})-\Sigma_{z}H_{E}(\bar{\bm{k}})\Sigma_{z}H(\bar{\bm{k}}))$, respectively. 
The matrix $T(\bar{\bm{k}})$ is a paraunitary matrix satisfying $T(\bar{\bm{k}})^{\dagger} \Sigma_{z} T(\bar{\bm{k}})=\Sigma_{z}$, which diagonalizes the magnon BdG Hamiltonian. 
Here, we introduce the phenomenological damping rate $1/\tau_{\ell}$ 
to take account of the finite lifetime of magnons.

\begin{figure}[H]
\centering
  \includegraphics[width=7cm]{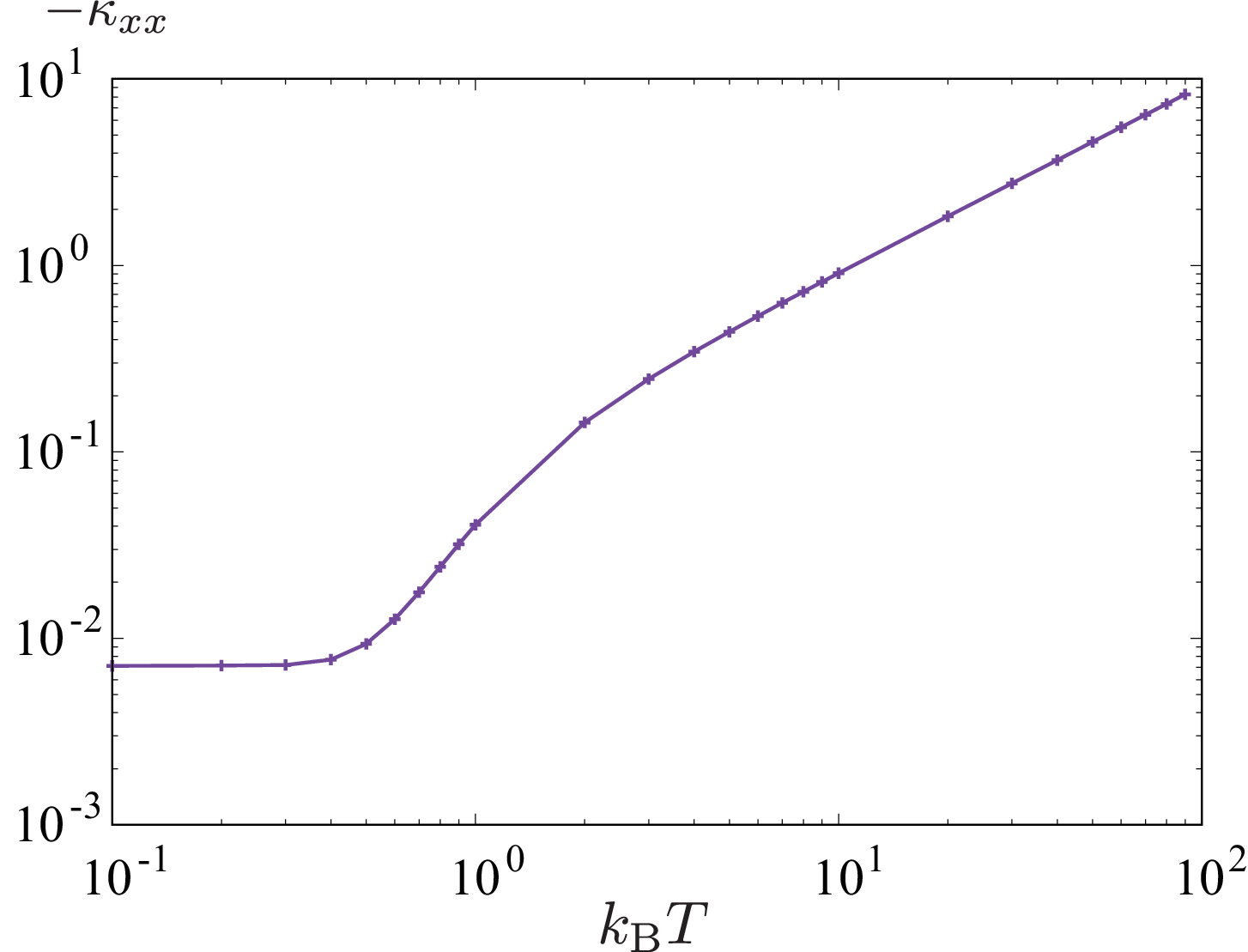}
\caption{(color online). Energy conductivity of the model (\ref{eq:Ham_AA}) as a function of temperature.
The data are calculated 
for $1/\tau_{\ell}=10^{-2}$ and $N=24$.
We take the $\bar{\bm{k}}$ summation over $80 \times 80$ grid points in the Brillouin zone.
The parameters are the same as those in Fig.~\ref{fig:HoneycombTCI_EF}.
}\label{fig:current}
\end{figure}

Figure \ref{fig:current} shows the conductivity 
$\kappa_{xx} = {\rm Re}\left[\langle J_{x} \rangle\right]/(g\mu_{\rm B}E_{x})$
as a function of temperature $T$.
We can see that the conductivity is a monotonically increasing function of temperature.
In the zero-temperature limit, $\kappa_{xx}$ should become zero as magnons cannot be excited,
whereas it appears to be nonzero.
This is due to a finite size effect.
We expect that the response discussed here is within the observable range because the AC phase due to an electric field has been observed in experiments, e.g., in a single-crystal yttrium iron garnet~\cite{Zhang14, comment_AC}.

\begin{figure}[H]
\centering
  \includegraphics[width=9cm]{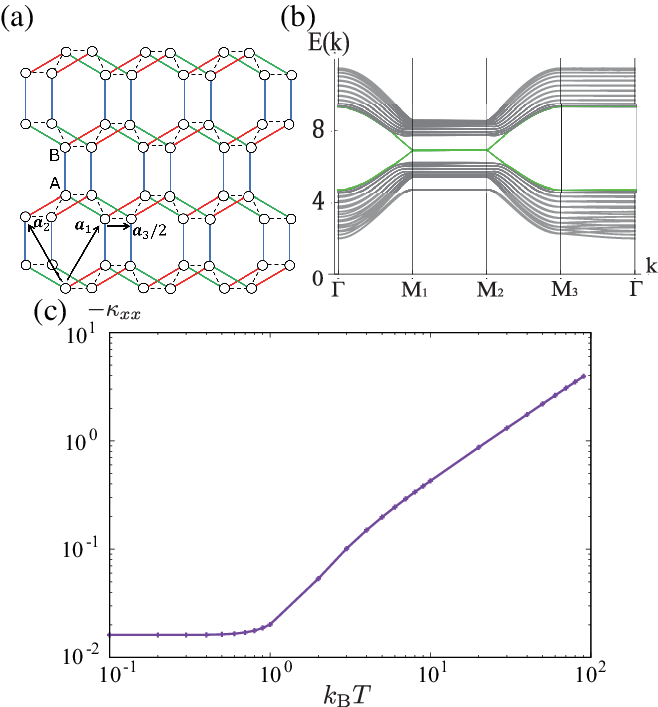}
\caption{(color online). (a) Monoclinic honeycomb-layer stacking structure of CrI$_3$.
Red, green, and blue bonds on the honeycomb lattice are the $x$-, $y$-, and $z$-bonds, respectively.
The black dashed lines represent the  couplings of the third term in Eq.~(\ref{eq:Ham_mono}).
The spins in odd and even layers point upward and downward, respectively, as in Fig.~\ref{fig:HoneycombTCI_AA}.
The lattice primitive vectors are written as $\bm{a}_{1}$, $\bm{a}_{2}$, and $\bm{a}_{3}$.
The arrow along the horizontal black dashed lines correspond to $\bm{a}_{3}/2$.
(b) The magnon band structure of a slab with $(100)$ face of the model~(\ref{eq:Ham_mono}). 
Topologically protected Dirac states are shown in green. 
(c) Energy conductivity of the model~(\ref{eq:Ham_mono}) as a function of temperature.
The parameters for (b) and (c) are 
$J=1.0$, $K=-0.5$, $\Gamma =0.3$, $D=0.07$, $J'=0.1, \kappa =0.4$, and $S=3/2$.
We set $1/\tau_{\ell}=10^{-2}$, $N=24$, and take the $\bar{\bm{k}}$ summation over $80 \times 80$ grid points in (c).
}\label{fig:HoneycombTCImono}
\end{figure}

{\it Another model: CrI$_{3}$.}---
Let us consider another model describing a magnetic compound CrI$_{3}$~\cite{McGuire15,Huang17,Huang18,Chen18,Sivadas18,Soriano19,Costa20,Soriano20,Ubrig19} and point out that it is a candidate material for MAFTI.
Here, CrI$_{3}$ is a van der Waals material in which the magnetic moments are carried by Cr$^{3+}$ ions with electronic configuration $3d^3$ forming a honeycomb lattice structure.
The spin magnitude of each Cr$^{3+}$ ion is $S=3/2$.
The crystal structure of the stacked honeycomb lattice magnet CrI$_{3}$ is monoclinic (rhombohedral) at a high (low) temperature.
The Hamiltonian of CrI$_{3}$ with the monoclinic structure illustrated in Fig.~\ref{fig:HoneycombTCImono}(a) is given by 
\begin{align}
\mathcal{H}=
&\sum_{l} \sum_{\gamma=x,y,z} \sum_{\langle ij \rangle_\gamma}H_{ij,l}^{\gamma}
+D\sum_{\langle\langle ij \rangle\rangle,l}\xi_{ij}\left(\bm{S}_{i,l}\times\bm{S}_{j,l}\right)_{z}\nonumber \\
&+J'\!\!\!\!\sum_{\langle (i,l),(j,l') \rangle\in {\rm mono}}\!\!\!\!\bm{S}_{i,l}\cdot\bm{S}_{j,l'} -\kappa \sum_{i}(S_{i}^{z})^{2}.
\label{eq:Ham_mono}
\end{align}
Here, $\langle ij \rangle_\gamma$ ($\gamma=x,$ $y,$ $z$) denotes a pair of the nearest neighbor sites $i$ and $j$ on the $x$-, $y$-, and $z$-bonds
which are colored with red, green, and blue in Fig.~\ref{fig:HoneycombTCImono}(a), respectively. 
The contribution from the $z$-bond in the $l$th layer is written as 
\begin{align}
H_{ij,l}^{z}=-J\bm{S}_{i,l}\!\cdot\!\bm{S}_{j,l}+K S_{i,l}^{z}S_{j,l}^{z}\!+\!\Gamma( S_{i,l}^{x}S_{j,l}^{y}\!+\!S_{i,l}^{y}S_{j,l}^{x}).
\label{eq:zbond}
\end{align}
The first and second terms in Eq.~(\ref{eq:zbond}) are the ferromagnetic Heisenberg and Kitaev interactions, respectively.
The third term is the symmetric off-diagonal intralayer interaction.
We can obtain the contributions from the $x$- and $y$-bonds ($H_{ij,l}^{x}$ and $H_{ij,l}^{y}$) by a cyclic permutation among $S^{x}$, $S^{y}$, and $S^{z}$.
The second term in Eq.~(\ref{eq:Ham_mono}) represents the DM interaction between intralayer next-nearest neighbor sites.
The remaining terms are the antiferromagnetic Heisenberg interaction between the nearest-neighbor layers and the easy axis anisotropy.
The summation $\sum_{\langle (i,l),(j,l') \rangle\in {\rm mono}}$ is taken over the nearest neighbor bonds across the layers in the monoclinic structure, which are shown in the black dashed lines in Fig.~\ref{fig:HoneycombTCImono}(a).
The magnon Hamiltonian of the model is given in Supplemental materials~\cite{comment_Suppl}.

Figure~\ref{fig:HoneycombTCImono}(b) shows the band structure for the model (\ref{eq:Ham_mono}) of a slab with $(100)$ face.
We can find a single Dirac cone at ${\rm M}_{2}$ point while the surface states are nearly degenerate from ${\rm M}_{1}$ to ${\rm M}_{2}$.
Here, we used the parameters estimated by density functional theory calculations for CrI$_3$~\cite{Soriano19, Kvashnin20}.
Since the band split between ${\rm M}_{1}$ and ${\rm M}_{2}$ 
comes from the $\Gamma$ interaction which breaks spin conservation and the interlayer coupling, we can see a more distinct single Dirac cone when $\Gamma$ and $J'$ are larger.
The surface states are protected by $S$-symmetry, and then they shift upward and downward under an electric field
via the spin-momentum locking, as in Fig.~\ref{fig:HoneycombTCI_EF}(b).
As discussed in the previous section, such a system exhibits the energy current induced by an electric field.
Figure~\ref{fig:HoneycombTCImono}(c) shows the energy conductivity $\kappa_{xx}$ as a function of temperature $T$.
The behavior is similar to that in the model~(\ref{eq:Ham_AA}) [See Fig.~\ref{fig:current}], while the starting points of the increase are different due to the difference between the 
heights of the magnon bands.
We also confirm that such a topological phase appears in a wide range of parameters by constructing a phase diagram of the model (\ref{eq:Ham_mono})~\cite{comment_Suppl}.


{\it Summary.}---
In this Letter, we have constructed a model for a magnet which has surface states of magnons protected by the combined symmetry of time-reversal and half translation.
The single Dirac surface states of the system appear as in the strong topological insulators in class AII, and thus it is expected to be robust against disorder as long as the respected symmetry is preserved~\cite{comments}.
We have also demonstrated that an electric field shifts the position of the Dirac cones in one and the other surfaces oppositely, resulting in an energy current. 
So far, the AC effect of magnons in a homogeneous electric field has not attracted much attention since the outcome is merely the shift of the wave vector in most cases.
However, our study has revealed that nontrivial response in magnon systems to a homogeneous electric field could be realized by the presence of topologically protected surface states and the interactions which break conservation of $S_{z}$.
A promising candidate material which exhibits this physics could be CrI$_{3}$ with a monoclinic structure.
Although the structure is realized at a temperature higher than 200K, 
it has been reported that the monoclinic structure remains unchanged in a thin film of CrI$_{3}$ even when the temperature is lowered below 200K~\cite{Ubrig19}.
We expect the magnon physics discussed in this Letter can be realized in a thin film of CrI$_{3}$, and similar materials.
Observation of the (shift of) Dirac surface states and magnon current under an electric field would provide the smoking-gun evidence for the realization of the topological phase, MAFTI.


{\it Acknowledgements.}---
We thank Hosho Katsura for valuable discussions and helpful comments on the manuscript.
This work was supported by JSPS KAKENHI Grants Nos. JP17K14352, JP20K14411, JP20J12861
 and JSPS Grant-in-Aid for Scientific Research on Innovative Areas ``Topological Materials Science'' (KAKENHI Grants No. JP18H04220) and ``Quantum Liquid Crystals'' (KAKENHI Grants No. JP20H05154).
H. K. was supported by the  JSPS through Program for Leading Graduate Schools (ALPS).
Y. A. also thanks the Okinawa Institute of Science and Technology Graduate University for the use of the facilities, Deigo cluster.

\begin{widetext}
\begin{center}
\Large{Supplemental materials for: Dirac surface states in magnonic analogs of topological crystalline insulators}
\end{center}

\begin{center}
\bf{
Explicit expression of Hamiltonian matrix in Eq.~(2)}
\end{center}

In this part, we show the explicit expression of the matrix $H(\bm{k})$ in the Hamiltonian~(2). 
The Hamiltonian of noninteracting bosonic systems is generally given by 
\begin{align}
H(\bm{k})=
\left(
\begin{array}{cc}
h(\bm{k}) &\Delta^{\dagger}(\bm{k}) \\
\Delta(\bm{k})&h^{*}(\bm{-k}) \\
\end{array} 
\right).
\tag{S.1}
\label{eq:Ham_magnon}
\end{align}
In the model~(2),
the $4 \times 4$ matrices $h(\bm{k})$ and $\Delta(\bm{k})$ are written as follows:
\begin{align}
&h(\bm{k})=
\left(
\begin{array}{cccc}
d+p(\bm{k})  &-\gamma_{+}(\bm{k}) &0 &0  \\
-\gamma^{*}_{+}(\bm{k}) &d-p(\bm{k}) &0 &0  \\
0 &0 &d-p(\bm{k}) &-\gamma_{+}(\bm{k})  \\
0 &0 &-\gamma^{*}_{+}(\bm{k}) &d+p(\bm{k})  \\
\end{array} 
\right), 
\tag{S.2}
\label{eq:h}\\
&\Delta(\bm{k})=
\left(
\begin{array}{cccc}
0  &-\gamma_{-}(\bm{k}) &\gamma^{*}_{z}(k_{3}) &0  \\
-\gamma^{*}_{-}(\bm{k}) &0 &0 &\gamma^{*}_{z}(k_{3})  \\
\gamma_{z}(k_{3}) &0 &0 &-\gamma_{-}(\bm{k})  \\
0 &\gamma_{z}(k_{3}) &-\gamma^{*}_{-}(\bm{k}) &0  \\
\end{array} 
\right),
\tag{S.3}
\label{eq:Delta}
\end{align}
where
\begin{align}
&d=J^{z}_{0}S+J^{z}_{1}S+J^{z}_{2}S+2J'S,  \tag{S.4}\\
&p(\bm{k})=2DS\left[\sin(k_{1})-\sin(k_{2})-\sin(k_{1}-k_{2})\right], \tag{S.5}\\
&\gamma_{\pm}(\bm{k})=J^{\pm}_{0}S+J^{\pm}_{1}Se^{ik_{1}}+J^{\pm}_{2}Se^{ik_{2}}, \tag{S.6}\\
&\gamma_{z}(k_{3})=J'S\left(1+e^{ik_{3}}\right).  \tag{S.7}
\end{align}
Here $S$, $k_i$, and $J^{\pm}_{i}$ are the spin magnitude,
$k_{i}=\bm{k}\cdot \bm{a}_{i}$,
and $J^{\pm}_{n}=(J^{x}_{n}\pm J^{y}_{n})/2$, respectively.

\begin{center}
\bf{Lattice deformation}
\end{center}


In the main text, we computed the magnon band structures and the energy conductivities in the models (1) and (12) by deforming the stacked honeycomb lattice into a topologically equivalent cubic-shaped lattice for simplicity.
We here explain the detail of the deformation.
Figure~\ref{fig:deformed}(a) shows the deformed honeycomb lattice of each layer.
When we take the lattice primitive vectors $\bm{a}_{1}=(1,0,0)$,  $\bm{a}_{2}=(0,1,0)$, and $\bm{a}_{3}=(0,0,1)$ as shown in the figure, the unit cell becomes cubic.
We note that layers are located at intervals of $\bm{a}_{3}/2$ so that the magnetic unit cell is cubic because the spins on odd and even layers are aligned in the opposite directions in the systems [See also Fig.~1 in the main text].
Figure~1(b) shows the Brillouin zone under open (periodic) boundary conditions in the $x$-direction ($y$- and $z$-directions).
The symmetry points $\Gamma$, ${\rm M}_1$, ${\rm M}_2$, and ${\rm M}_3$ correspond to those in Figs.~2 and 4(b) in the main text.

\begin{figure}[H]
\vspace{18mm}
\centering
  \includegraphics[width=14cm]{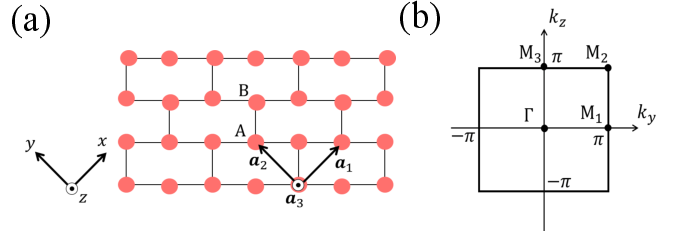}
\vspace{-8mm}
\caption{(color online). (a) Deformed honeycomb lattice, which is a layer of the stacked honeycomb lattice. 
The lattice vectors are taken to be $\bm{a}_{1}=(1,0,0)$, $\bm{a}_{2}=(0,1,0)$, and $\bm{a}_{3}=(0,0,1)$. 
(b) The Brillouin Zone in the case 
of open (periodic) boundary conditions in the $\bm{a}_{1}$ ($\bm{a}_{2}$ and $\bm{a}_{3}$) direction(s). The high-symmetry points $(k_{2},k_{3})=(0,0),$ $(\pi,0),$ $(\pi,\pi),$ and $(\pi,0)$ are indicated as $\Gamma$, ${\rm M}_1$, ${\rm M}_2$, and ${\rm M}_3$, respectively.
}\label{fig:deformed}
\end{figure}

\begin{center}
\bf{Expression of perturbation term}
\end{center}
In the main text, we use the linear response theory to evaluate the energy current induced by an electric field $\bm{E}=E_{x}e^{-i\omega t}\bm{e}_{x}$.
In this part, we show the expression of the perturbation Hamiltonian $H_E$.
By taking account of the AC effect, 
we can describe the magnon Hamiltonian as follows:
\begin{align}
H=\sum_{ij}\left(h_{ij}(E_{x})b_{i}^{\dagger}b_{j}+\Delta_{ij}(E_{x})b_{i}b_{j}\right)+h.c..
\tag{S.8}
\end{align}
Then, the perturbation Hamiltonian in the first order of  the electric field $E_x$
can be expressed as $H_{E}(t)=H_{E}e^{-i\omega t}$, where $H_{E}$ is defined as
\begin{align}
H_{E}=\sum_{ij}\left(E_{x}\left[\frac{\partial}{\partial E_{x}}h_{ij}(E_{x})
\right]_{E_{x}=0}b_{i}^{\dagger}b_{j}+ E_{x}\left[\frac{\partial}{\partial E_{x}}\Delta_{ij}(E_{x})
\right]_{E_{x}=0}b_{i}b_{j}\right) +h.c..
\tag{S.9}
\label{eq:HE_real}
\end{align}
Here, we assume the open (periodic) boundary conditions in the direction(s) along $\bm{a}_{1}$ ($\bm{a}_{2}$ and $\bm{a}_{3}$).
By applying the Holstein-Primakoff and Fourier transformations to the Hamiltonian (\ref{eq:HE_real}), we obtain the following expression:
\begin{align}
H_{E}=\frac{1}{2}\sum_{\bar{\bm{k}}}\bm{\psi}^{\dagger}(\bar{\bm{k}})H_{E}(\bar{\bm{k}})\bm{\psi}(\bar{\bm{k}}),
\tag{S.10}
\label{eq:op_mat}
\end{align}
where $\bm{\psi}(\bar{\bm{k}})$ is defined as
\begin{align}
\bm{\psi}(\bar{\bm{k}})=(\bm{b}(\bar{\bm{k}}),\bm{b}^{\dagger}(-\bar{\bm{k}}))^{T}.
\tag{S.11}
\label{eq:psi}
\end{align}
The operator $\bm{b}(\bar{\bm{k}})$ consists of $4N$ components when the number of the unit cells in the $x$-direction is $N$.
Each component is given as follows:
\begin{align}
b_{4n+1}(\bar{\bm{k}})=b_{n}(\bar{\bm{k}},A,1), \nonumber \\
b_{4n+2}(\bar{\bm{k}})=b_{n}(\bar{\bm{k}},B,1), \nonumber \\
b_{4n+3}(\bar{\bm{k}})=b_{n}(\bar{\bm{k}},A,2), \nonumber \\
b_{4n+4}(\bar{\bm{k}})=b_{n}(\bar{\bm{k}},B,2). 
\tag{S.12}
\label{eq:b}
\end{align}
The operator $b_{j}(\bar{\bm{k}},A(B),1(2))$ annihilates a magnon at the sublattice $A(B)$ in the $j$th unit cell on the layer with odd (even) $l$.
The matrix $H_{E}(\bar{\bm{k}})$ in Eq.~(\ref{eq:op_mat}) can be written as follows:
\begin{align}
&H_{E}(\bar{\bm{k}})=
\left(
\begin{array}{cccc}
 h_{E}(\bar{\bm{k}}) &\Delta^{\dagger}_{E}(\bar{\bm{k}})  \\
 \Delta_{E}(\bar{\bm{k}}) &h^{*}_{E}(-\bar{\bm{k}})  \\
\end{array} 
\right),
\tag{S.13}
\label{eq:H_E}
\end{align}
where the matrices $h_{E}(\bar{\bm{k}})$ and $\Delta_{E}(\bar{\bm{k}})$ are given by
\begin{align}
&h_{E}(\bar{\bm{k}})=
\left(
\begin{array}{cccc}
h_{E1}(\bar{\bm{k}}) &h_{E2}(\bar{\bm{k}})&0 &0  \\
h^{\dagger}_{E2}(\bar{\bm{k}}) &\ddots  &\ddots &0  \\
0 &\ddots &\ddots &h_{E2}(\bar{\bm{k}})  \\
0 &0 &h^{\dagger}_{E2}(\bar{\bm{k}}) &h_{E1}(\bar{\bm{k}})   \\
\end{array} 
\right), 
\tag{S.14}
\\
&\Delta_{E}(\bar{\bm{k}})=
\left(
\begin{array}{cccc}
\Delta_{E1}(\bar{\bm{k}}) &\Delta_{E2}(\bar{\bm{k}})&0 &0  \\
\Delta_{E2}(-\bar{\bm{k}}) &\ddots  &\ddots &0  \\
0 &\ddots &\ddots &\Delta_{E2}(\bar{\bm{k}})  \\
0 &0 &\Delta_{E2}(-\bar{\bm{k}}) &\Delta_{E1}(\bar{\bm{k}})   \\
\end{array} 
\right).
\tag{S.15}
\end{align}
Here, the matrices  $h_{E1}(\bar{\bm{k}})$, $h_{E2}(\bar{\bm{k}})$, $\Delta_{E1}(\bar{\bm{k}})$, and $\Delta_{E2}(\bar{\bm{k}})$  for the model~(1)  are defined as 
\begin{align}
&h_{E1}(\bar{\bm{k}})=S \left(
\begin{array}{cccc}
2D \cos(k_{2}) &i J^{+}_{2}e^{ik_{2}} &0 &0  \\
-i J^{+}_{2}e^{-ik_{2}} &-2D \cos(k_{2})  &0 &0  \\
0 &0 &-2D \cos(k_{2})  &-i J^{+}_{2}e^{ik_{2}}  \\
0 &0 &i J^{+}_{2}e^{-ik_{2}} &2D \cos(k_{2})   \\
\end{array} 
\right) ,
\tag{S.16}
\\
&h_{E2}(\bar{\bm{k}})
=S\left(
\begin{array}{cccc}
-D(1+e^{-ik_{2}}) &\frac{i}{2}J^{+}_{1} &0 &0 \\
0 &D(1+e^{-ik_{2}}) &0 &0 \\
0 &0 &-D(1+e^{-ik_{2}}) &-\frac{i}{2}J^{+}_{1} \\
0 &0 &0 &D(1+e^{-ik_{2}}), \\
\end{array} 
\right). 
\tag{S.17}
\\
&\Delta_{E1}(\bar{\bm{k}}) =0, 
\tag{S.18}
\\
&\Delta_{E2}(\bar{\bm{k}}) =0.
\tag{S.19}
\end{align}
In the case of the model~(12) for CrI$_3$, these matrices are given by
\begin{align}
&h_{E1}(\bar{\bm{k}})=S \left(
\begin{array}{cccc}
2D \cos(k_{2}) &i J e^{ik_{2}} &0 &0  \\
-i J e^{-ik_{2}} &-2D \cos(k_{2})  &0 &0  \\
0 &0 &-2D \cos(k_{2})  &-i J e^{ik_{2}}  \\
0 &0 &i J e^{-ik_{2}} &2D \cos(k_{2})   \\
\end{array} 
\right), 
\tag{S.20}
\\
&h_{E2}(\bar{\bm{k}}) =S \left(
\begin{array}{cccc}
-D(1+e^{-ik_{2}}) &\frac{i}{2}J +\frac{i}{4}K  &0 &0 \\
0 &D(1+e^{-ik_{2}}) &0 &0 \\
0 &0 &-D(1+e^{-ik_{2}}) &-\frac{i}{2}J-\frac{i}{4}K  \\
0 &0 &0 &D(1+e^{-ik_{2}}) \\
\end{array} 
\right),  
\tag{S.21}
\\
&\Delta_{E1}(\bar{\bm{k}}) =S\left(
\begin{array}{cccc}
0 &0 &\frac{i}{6}J' (- 1+e^{-i k_{3}}) &-\frac{i}{3}J' e^{-ik_{2}} \\
0 &0 &\frac{i}{6}J' e^{i(k_{2}-k_{3})} &\frac{i}{6}J' (-1+e^{-i k_{3}}) \\
\frac{i}{6}J' (-1+e^{i k_{3}}) &\frac{i}{6}J' e^{i(-k_{2}+k_{3})} &0 &0 \\
-\frac{i}{3}J' e^{ik_{2}} &\frac{i}{6}J' (-1+e^{i k_{3}}) &0 &0 \\
\end{array} 
\right),  
\tag{S.22}
\\
&\Delta_{E2}(\bar{\bm{k}})=S\left(
\begin{array}{cccc}
0 &0 &0 &0  \\
0 &0 &\frac{i}{3}J' &0  \\
0 &0 &0 &0  \\
-\frac{i}{6}J'e^{ik_{3}} &0 &0 &0  \\
\end{array} 
\right).
\tag{S.23}
\end{align}

\begin{center}
\bf{Expression of energy current}
\end{center}
In this part, we provide a detailed expression of the energy current operator $J_{x}$.
Firstly, we assume the same boundary conditions as in the previous part, i.e.,
the open (periodic) boundary conditions in the direction(s) along $\bm{a}_{1}$ 
($\bm{a}_{2}$ and $\bm{a}_{3}$). 
By applying the Holstein-Primakoff transformation and Fourier transformation, we can 
rewrite the Hamiltonian~(1) as follows:
\begin{align}
H=\frac{1}{2}\sum_{\bar{\bm{k}}}\bm{\psi}^{\dagger}(\bar{\bm{k}})H(\bar{\bm{k}})\bm{\psi}(\bar{\bm{k}})
\tag{S.24}
\label{eq:op_mat2}.
\end{align}
Here, the definition of $\bm{\psi}(\bar{\bm{k}})$ is given by Eqs.~(\ref{eq:psi}) and (\ref{eq:b}).
Then, the position operator can be written as
\begin{align}
P=\frac{1}{2}\sum_{\bar{\bm{k}}}\bm{\psi}^{\dagger}(\bar{\bm{k}})\bar{P}\bm{\psi}(\bar{\bm{k}}),
\tag{S.25}
\end{align}
where $\bar{P}$ is an $8N\times 8N$ diagonal matrix.
Each components of $\bar{P}$ is written as
\begin{align}
\bar{P}_{4n+1,4n+1}=\bar{P}_{4n+3,4n+3}=\bar{P}_{4(n+N)+1,4(n+N)+1}=\bar{P}_{4(n+N)+3,4(n+N)+3}=\frac{\sqrt{3}}{2}(n-1)-\frac{1}{2}\left( \frac{\sqrt{3}}{2}(N-1)+\frac{1}{2\sqrt{3}}\right),
\tag{S.26} \\
\bar{P}_{4n+2,4n+2}=\bar{P}_{4n+4,4n+4}=\bar{P}_{4(n+N)+2,4(n+N)+2}=\bar{P}_{4(n+N)+4,4(n+N)+4}=\frac{\sqrt{3}}{2}(n-1)-\frac{1}{2}\left( \frac{\sqrt{3}}{2}(N-1)-\frac{1}{2\sqrt{3}}\right).
\tag{S.27}
\end{align}
By using the operators $\bar{P}$ and $H(\bar{\bm{k}})$, we define the energy current operator $J_{x}$ such that it satisfies the continuity equation in the following: 
\begin{align}
J_{x}=\sum_{i}\bar{P}_{i}\dot{h}_{i},
\tag{S.28}
\end{align}
where $\dot{h}_{i}$ is the time derivative of the Hamiltonian density at the site $i$ and is defined as 
\begin{align}
&\dot{h}_{i}=i [H, h_{i}],  
\tag{S.29}
 \\
&h_{i}=\frac{1}{2}\sum_{\bar{\bm{k}}}\sum_{j}\psi_{i}^{\dagger}(\bar{\bm{k}})H_{ij}(\bar{\bm{k}})\psi_{j}(\bar{\bm{k}}) .
\tag{S.30}
\end{align}
By further calculation, we obtain the energy current operator $J_{x}$ as follows:
\begin{align}
J_{x}=-\frac{i}{4}\sum_{\bar{\bm{k}}}\bm{\psi}^{\dagger}(\bar{\bm{k}})
\left(\bar{P}H(\bar{\bm{k}})\Sigma_{z}H(\bar{\bm{k}})-H(\bar{\bm{k}})\Sigma_{z}H(\bar{\bm{k}})\bar{P}\right)
\bm{\psi}(\bar{\bm{k}}),
\tag{S.31}
\end{align}
where $\Sigma_{z}=\sigma_{z}\otimes 1_{4N}$.
Then, we define a matrix $J_{x}(\bar{\bm{k}})$ as 
\begin{align}
J_{x}(\bar{\bm{k}})=
-\frac{i}{2}\left(\bar{P}H(\bar{\bm{k}})\Sigma_{z}H(\bar{\bm{k}})-H(\bar{\bm{k}})\Sigma_{z}H(\bar{\bm{k}})\bar{P}\right).
\tag{S.32}
\end{align}

\begin{center}
\bf{Magnon Hamiltonian of CrI$_3$}
\end{center}
In the main text, we point out that CrI$_3$ is a candidate material for MAFTI, whose spin Hamiltonian is expressed by Eq.~(12).
In this part, we show the magnon Hamiltonian for CrI$_3$ in detail, which has the same form as Eq.~(\ref{eq:Ham_magnon}).
The matrices $h(\bm{k})$ and $\Delta(\bm{k})$ in this system are given by
\begin{align}
&h(\bm{k})=
\left(
\begin{array}{cccc}
d+p(\bm{k})  &-\gamma(\bm{k})+\frac{KS}{2}(1-e^{i k_{1}}) &0 &0  \\
-\gamma^{*}(\bm{k})+\frac{KS}{2}(1-e^{-i k_{1}})  &d-p(\bm{k}) &0 &0  \\
0 &0 &d-p(\bm{k}) &-\gamma(\bm{k})+\frac{KS}{2}(1-e^{i k_{1}})  \\
0 &0 &-\gamma^{*}(\bm{k})+\frac{KS}{2}(1-e^{-i k_{1}}) &d+p(\bm{k})  \\
\end{array} 
\right), 
\tag{S.33}
\\
&\Delta(\bm{k})=
\left(
\begin{array}{cccc}
0  &\frac{KS}{2}(1+e^{i k_{1}})-i \Gamma S e^{i k_{2}} &\gamma^{*}_{3}(k_{3}) &\gamma_{21}^{*}(\bm{k})  \\
\frac{KS}{2}(1+e^{-i k_{1}})-i \Gamma S e^{-i k_{2}}  &0&\gamma_{12}^{*}(\bm{k})  &\gamma^{*}_{3}(k_{3})  \\
\gamma_{3}(k_{3}) &\gamma_{12}(\bm{k}) &0 &\frac{KS}{2}(1+e^{i k_{1}})+i \Gamma S e^{i k_{2}}   \\
\gamma_{21}(\bm{k}) &\gamma_{3}(k_{3}) &\frac{KS}{2}(1+e^{-i k_{1}})+i \Gamma S e^{-i k_{2}}  &0  \\
\end{array} 
\right),
\tag{S.34}
\end{align}
where
\begin{align}
&d=3JS+4J'S-KS, \tag{S.35}\\
&p(\bm{k})=2DS\left[\sin(k_{1})-\sin(k_{2})-\sin(k_{1}-k_{2})\right], \tag{S.36}\\
&\gamma(\bm{k})=JS\left(1+e^{ik_{1}}+e^{ik_{2}}\right), \tag{S.37}\\
&\gamma_{3}(k_{3})=J'S\left(1+e^{ik_{3}}\right),  \tag{S.38}  \\
&\gamma_{21}(\bm{k})=J'S\left(e^{ik_{2}}+e^{i (k_{1}+k_{3})}\right),   \tag{S.39}\\
&\gamma_{12}(\bm{k})=J'S\left(e^{-ik_{1}}+e^{i (-k_{2}+k_{3})}\right). \tag{S.40}
\end{align}

\begin{center}
\bf{Phase diagrams}
\end{center}

In this part, we construct  phase diagrams of models (1) and (12) in the main text which we discussed in the main text.
Figure~\ref{fig:phasediagram}(a) shows a phase diagram 
in the model~(1)
as a function of 
 $D$ and $\alpha$ which is a parameter introduced as $J_{0}^{x}=1.0+\alpha$ and $J_{2}^{x}=1.0-\alpha$. 
Figure~\ref{fig:phasediagram}(b) is the one 
in the model~(12) as a function of $K$ and $D$.
The ``Strong'' (``Weak'') topological phase 
 with a slab structure has surface states of magnons between the two bulk bands, which 
cross at only ${\rm M}_{2}$ point (both ${\rm M}_{1}$ and ${\rm M}_{2}$ points).
The single Dirac gapless states in the ``Strong''  topological phase are protected by $S$-symmetry and robust against disorder, while the gapless states in the ``Weak'' topological phase are not.
What we mention as the ``Strong" topological phase here is nothing but MAFTI in the main text.
The ``Trivial''  topological phase does not have gapless surface states.
When the gap between the two bulk bands 
is closed, 
we call it ``Gapless'' phase.
In particular, the system becomes a magnonic analog of nodal line semimetals when the DM interaction is zero.

\begin{figure}[H]
\centering
\vspace{18mm}
  \includegraphics[width=16cm]{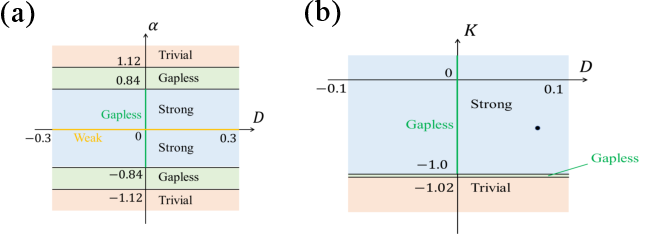}
\vspace{-8mm}
\caption{(color online). (a) Phase diagram of 
the model~(1).
The parameters are chozen to be $J_{0}^{x}=1.0+\alpha$,  $J_{1}^{x}=1.0$, $J_{2}^{x}=1.0-\alpha$, $J_{0}^{y}=J_{1}^{y}=J_{2}^{y}=1.0$,  $J_{0}^{z}=J_{1}^{z}=J_{2}^{z}=1.4$, $J'=0.5$, and $S=1.0$.
 (b) Phase diagram of 
 the model~(12). The other parameters are chozen to be $J=1.0$ , $\Gamma =0.3$, $J'=0.1, \kappa =0.4$, and $S=3/2$. The black dot with $D=0.07$ and $K=-0.5$ corresponds to the set of parameters which is used in Fig.~4 in the main text.
}\label{fig:phasediagram}
\end{figure}

\end{widetext}


\begin{thebibliography}{120}
\bibitem{comment_artificial} 
In the models for the magnonic analog of topological insulators in three dimensions~\cite{Kondo19b}, two spins are localized at each site of the diamond lattice and point in opposite directions.
In addition, fine-tuning of quite complicated interactions is required to realize the topological phase.
\bibitem{comment_role} The DM interaction plays a role in making the magnon bands topologically nontrivial.
The XYZ anisotropy, which 
breaks spin conservation, and bond dependence of the Heisenberg interaction 
lift the degeneracy of surface states from ${\rm M}_{1}$ to ${\rm M}_{2}$ points, resulting in a single Dirac cone at ${\rm M}_{2}$ point.
It 
is also noted that without the interlayer interaction, the system would be 
merely stacked Chern insulators of magnons with alternating Chern numbers.
\bibitem{comment_AC}
The AC phase of magnons 
can also be understood 
in terms of electric-field-induced DM interaction (via the spin-orbit interaction) due to the superexchange mechanism~\cite{Liu-Vignale11, Katsura05}, which is proportional to the strength of the electric field.
In fact, the form of 
the phase 
is identical to that of the AC phase, and thus
they are indistinguishable 
from each other.
The spin-orbit interaction itself  induced by an electric field  in a vacuum is quite small.
However, the interaction can be drastically enhanced in solids as we observed a variety of phenomena originating from the spin-orbit interaction.
\end{thebibliography}
\end{document}